\documentclass[12pt,preprint]{aastex}
\usepackage{epsfig}
\usepackage{natbib}
\usepackage{graphicx}
\usepackage{slashbox}
\usepackage{multirow}
\usepackage{lscape}
\usepackage{mathrsfs,amssymb}
\usepackage{amsmath}
\newcommand       \Angstrom     {\,{\rm \AA}}

\newcommand       \cm           {\,{\rm cm}}

\newcommand       \K            {\,{\rm K}}

\newcommand       \simlt        {\lesssim}

\newcommand       \mum          {\,{\rm \mu m}}

\newcommand       \Teff         {T_{\star}}

\newcommand       \simali       {\sim\,}

\newcommand       \Iaro         {I_{3.3}}
\newcommand       \Iali         {I_{3.4}}

\newcommand       \Aratio        {A_{3.4}/A_{3.3}}
\newcommand       \Aali           {A_{3.4}}
\newcommand       \Aaro          {A_{3.3}}
\newcommand       \Adfa       {A_{6.85}}

\newcommand       \Adfb       {A_{7.25}}

\newcommand       \Acc        {A_{6.2}}

\newcommand       \NC         {N_{\rm C}}
\newcommand       \NCaro      {N_{\rm C,aro}}
\newcommand       \NCali      {N_{\rm C,ali}}
\newcommand       \nuMea         {\nu_{\rm Me,1}}
\newcommand       \nuMeb         {\nu_{\rm Me,2}}
\newcommand       \nuMec         {\nu_{\rm Me,3}}
\newcommand       \km        {\,{\rm km}}
\newcommand       \mol       {\,{\rm mol}}

\newcommand       \Ali        {3.4}
\newcommand       \Aro        {3.3}
\newcommand       \Arel       {A_{\rm ali}/A_{\rm aro}}

\newcommand       \fali      {f_{\rm ali}}
\newcommand       \faro      {f_{\rm aro}}
\newcommand       \Li        {L_i}
\newcommand       \Lj        {L_j}

\newcommand       \CabsPAH    {C^{\scriptscriptstyle\rm PAH}_{\rm abs}}
\newcommand       \Iratiomod   {\left(I_{3.4}/I_{3.3}\right)_{\rm mod}}
\newcommand \Iratioobs {\left(I_{3.4}/I_{3.3}\right)_{\rm obs}}
\newcommand \Iratiomodp   {\left(I_{3.4}/I_{3.3}\right)_{\rm mod}^{\prime}}
\countdef\decade=200
\advance\decade by \year
\countdef\hours=201
\advance\hours by \time
\divide\hours by 60
\countdef\mins=202
\advance\mins by \hours
\multiply\mins by 60
\multiply\hours by 100
\countdef\miltime=203
\advance\miltime by \hours
\advance\miltime by \time
\advance\miltime by -\mins

\shorttitle{Intensity Scaling for the C--H Stretching Modes}
\title{
\vspace*{-2.0em}
{\normalsize\rm accepted for publication in {\it The Astrophysical Journal}}\\
\vspace*{1.0em}
Polycyclic Aromatic Hydrocarbons with Aliphatic Sidegroups:
Intensity Scaling for the C--H Stretching Modes
and Astrophysical Implications
}
\author{X.J.~Yang\altaffilmark{1,2},
            Aigen Li\altaffilmark{2},
            R.~Glaser\altaffilmark{3},
            and J.X.~Zhong\altaffilmark{1}
            }
\altaffiltext{1}{Department of Physics,
                 Xiangtan University,
                 411105 Xiangtan, Hunan Province, China;
                 {\sf xjyang@xtu.edu.cn, jxzhong@xtu.edu.cn}
                 }
\altaffiltext{2}{Department of Physics and Astronomy,
                 University of Missouri,
                 Columbia, MO 65211, USA;
                 {\sf lia@missouri.edu}
                 }
\altaffiltext{3}{Department of Chemistry,
                 University of Missouri,
                 Columbia, MO 65211, USA;
                 {\sf glaserr@missouri.edu}
                 }

\begin{document}

\begin{abstract}
The so-called unidentified infrared emission (UIE) features
at 3.3, 6.2, 7.7, 8.6, and 11.3$\mum$
ubiquitously seen in a wide variety of astrophysical regions
are generally attributed to polycyclic aromatic hydrocarbon
(PAH) molecules.
Astronomical PAHs may have an aliphatic component
as revealed by the detection in many UIE sources of
the aliphatic C--H stretching feature at 3.4$\mum$.
The ratio of the observed intensity of
the 3.4$\mum$ feature
to that of the 3.3$\mum$ aromatic C--H feature
allows one to estimate
the aliphatic fraction
of the UIE carriers.
This requires the knowledge of
the intrinsic oscillator strengths
of the 3.3$\mum$ aromatic
C--H stretch ($\Aaro$) and the 3.4$\mum$ aliphatic
C--H stretch ($\Aali$).
Lacking experimental data
on $\Aaro$ and $\Aali$
for the UIE candidate materials,
one often has to rely on
quantum-chemical computations.
Although the second-order M$\o$ller-Plesset (MP2)
perturbation theory with a large basis set
is more accurate
than the B3LYP density functional theory,
MP2 is computationally very demanding
and impractical
for large molecules.
Based on methylated PAHs,
we show here that, by scaling
the band strengths computed
at an inexpensive level
(e.g., {\rm B3LYP/6-31G$^{\ast}$})
we are able to obtain band strengths
as accurate as that computed at
far more expensive levels
(e.g., MP2/{\rm 6-311+G(3df,3pd)}).
We calculate the model spectra
of methylated PAHs and their cations
excited by starlight of different spectral shapes
and intensities.
We find $\left(\Iali/\Iaro\right)_{\rm mod}$,
the ratio of the model intensity of
the 3.4$\mum$ feature
to that of the 3.3$\mum$ feature,
is insensitive to the spectral shape
and intensity of the exciting starlight.
We derive a straightforward relation 
for determining the aliphatic fraction 
of the UIE carriers (i.e., the ratio of
the number of C atoms in aliphatic units
$N_{\rm C,ali}$
to that in aromatic rings $N_{\rm C,aro}$)
from the observed band ratios
$\Iratioobs$:
$\NCali/\NCaro\approx0.57\times\Iratioobs$
for neutrals and
$\NCali/\NCaro\approx0.26\times\Iratioobs$
for cations.
\end{abstract}

\section{Introduction\label{sec:intro}}
The infrared (IR) spectra of a wide range of
galactic and extragalactic objects
with associated dust and gas
are dominated by a series of emission features
at 3.3, 6.2, 7.7, 8.6, 11.3, and 12.7$\mum$
(see Peeters 2014).
Collectively known as the ``unidentified'' IR
emission (IUE) features due to the fact that
the exact nature of their carriers remains
unknown (see Peeters et al.\ 2003, Yang et al.\ 2017),
the hypothesis of polycyclic aromatic hydrocarbon (PAH)
molecules as the carriers of the UIE features
has gained widespread acceptance and extreme popularity.
The PAH hypothesis attributes the UIE features
to the stretching and bending vibrational modes
of PAH molecules
(L\'eger \& Puget 1984, Allamandola et al.\ 1985).

While PAH is a precisely defined chemical term
(i.e., PAHs are fused benzene rings made up of
carbon and hydrogen atoms),
the PAH hypothesis does not really postulate
that astronomical PAHs are pure aromatic compounds
as strictly defined by chemists.
Instead, PAH molecules in astronomical
environments may include ring defects
(e.g., see Yu \& Nyman 2012),
substituents (e.g., N in place of C,
see Hudgins et al.\ 2005, Mattioda et al.\ 2008,
Alvaro Galu\'e et al.\ 2010,
Gruet et al.\ 2016, Gao et al.\ 2016;
O in place of C, see Bauschlicher 1998;
Fe in place of C, see Szczepanski et al.\ 2006,
Bauschlicher 2009, Simon \& Joblin 2010),
partial deuteration
(e.g., see Allamandola et al.\ 1989,
Hudgins et al.\ 2004, Peeters et al.\ 2004,
Draine 2006, Onaka et al.\ 2014),
partial dehydrogenation
(e.g., see Tielens et al.\ 1987,
Malloci et al.\ 2008)
and sometimes superhydrogenation
(e.g., see Bernstein et al.\ 1996,
Thrower et al.\ 2012, Sandford et al.\ 2013).
%

%
%
%

\begin{figure*}
\begin{center}
\includegraphics[scale=0.65,clip]{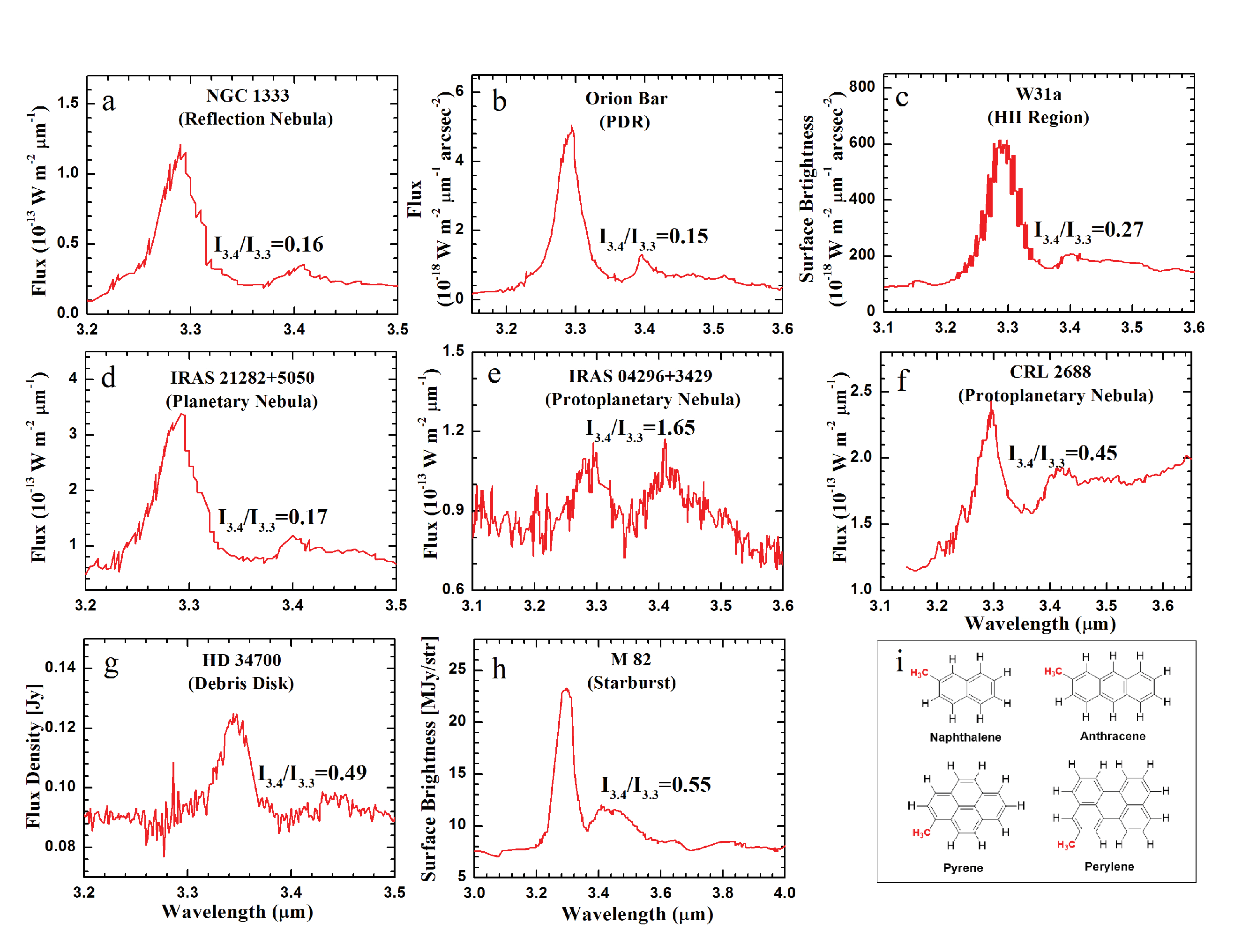}
\caption{\footnotesize
         \label{fig:Aro_Ali_Obs}
         Aromatic and aliphatic C--H stretching emission
         features seen in representative astronomical sources:
         (a) NGC\,1333 (reflection nebula, Joblin et al.\ 1996),
         (b) Orion Bar (photodissociated region [PDR],
              Sloan et al.\ 1997),
         (c) W31a (HII region, Mori et al.\ 2014),
         (d) IRAS\,21282+5050 (planetary nebula,
              Nagata et al.\ 1988),
         (e) IRAS\,04296+3429 (protoplanetary nebula,
              Geballe et al.\ 1992),
         (f) CRL\,2688 (protoplanetary nebula,
             Geballe et al.\ 1992),
         (g) HD\,34700 (debris disk,
              Smith et al.\ 2004),
         (h) M\,82 (starburst galaxy,
              Yamagishi et al.\ 2012), and
         (i) four methylated PAH molecules.
         }
\end{center}
\end{figure*}

Astronomical PAHs may likely also include
an aliphatic component, as revealed by
the detection in many UIE sources
of a weak satellite emission feature
at 3.4$\mum$ always which accompanies
the 3.3$\mum$ emission feature
(e.g., see Geballe et al.\ 1985, 1989,
Jourdain de Muizon et al.\ 1986, 1990,
Nagata et al.\ 1988,
Allamandola et al.\ 1989,
Sandford et al.\ 1991,
Joblin et al.\ 1996,
Sloan et al.\ 1997).
For illustration, we show in
Figure~\ref{fig:Aro_Ali_Obs}
the 3.3 and 3.4$\mum$ emission
features of several representative
astrophysical regions.
The 3.4$\mum$ feature is
generally thought to arise from
the C--H stretching vibration of
aliphatic hydrocarbon materials,
while the 3.3$\mum$ feature is due to
the C--H stretching mode of aromatic hydrocarbons.
Also detected in some UIE sources are
the aliphatic C--H deformation bands
at 6.85 and 7.25$\mum$
(see Sloan et al.\ 2014,
and see Table~3 in Yang et al.\ 2016a
for a summary).

In recent years, the aliphatic fraction
of the UIE carriers
--- the ratio of
the number of C atoms in aliphatic units
($N_{\rm C,ali}$)
to that in aromatic rings ($N_{\rm C,aro}$)
--- has received much attention
(e.g., see Kwok \& Zhang 2011,
Li \& Draine 2012, Rouill\'e et al.\ 2012,
Steglich et al.\ 2013, Yang et al.\ 2013, 2016a,b).
Kwok \& Zhang (2011) argued that the material
responsible for the UIE features may have
a substantial aliphatic component and therefore,
by definition, PAHs can not be the UIE carrier.
This argument can be tested by examining
the ratio of the observed intensity of
the 3.3$\mum$ feature ($\Iaro$)
to that of the 3.4$\mum$ feature ($\Iali$)
of UIE sources.
If the intrinsic oscillator strengths
(per chemical bond) of the 3.3$\mum$ aromatic
C--H stretch ($\Aaro$) and the 3.4$\mum$ aliphatic
C--H stretch ($\Aali$) are known,
one could drive the aliphatic fraction
of the UIE carriers from
$N_{\rm C,ali}/N_{\rm C,aro}\approx
0.3\times\,\left(I_{3.4}/I_{3.3}\right)
\times\,\left(A_{3.3}/A_{3.4}\right)$
(see Li \& Draine 2012).
Here the factor 0.3 arises from 
the assumption of one aliphatic C atom 
corresponding to 2.5 aliphatic C--H bonds 
(intermediate between methylene --CH$_2$
and methyl --CH$_3$) and one aromatic C atom 
corresponding to 0.75 aromatic C--H bond
(intermediate between benzene C$_6$H$_6$
and coronene C$_{24}$H$_{12}$).

Unfortunately, there is little experimental
information on $\Aaro$ and $\Aali$
for the UIE candidate materials.
Therefore, one often has to rely on
quantum-chemical computations
based on density functional theory
or second-order perturbation theory.
To this end, one often uses
the Gaussian09 software (Frisch et al.\ 2009)
and employs the hybrid density functional
theoretical method (B3LYP)
in conjunction with a variety of basis sets.
In the order of increasing accuracy
and computational demand,
the commonly adopted basis sets are
(see Pople et al.\ 1987):
{\rm 6-31G$^{\ast}$},
{\rm 6-31+G$^{\ast}$},
{\rm 6-311+G$^{\ast}$},
{\rm 6-311G$^{\ast\ast}$},
{\rm 6-31+G$^{\ast\ast}$},
{\rm 6-31++G$^{\ast\ast}$},
{\rm 6-311+G$^{\ast\ast}$},
{\rm 6-311++G$^{\ast\ast}$},
{\rm 6-311+G(3df,3pd)}, and
{\rm 6-311++G(3df,3pd)}.
One also often employs second-order
M$\o$ller-Plesset perturbation theory
(hereafter abbreviated as MP2)
in conjunction with these basis sets.
The MP2 method is thought to be more accurate
in computing band intensities than B3LYP
(see Cramer et al.\ 2004).
Indeed, as demonstrated in \S\ref{sec:B3LYP},
the IR intensities calculated
at the {\rm B3LYP/6-31G$^{\ast}$} level
for the 3.3$\mum$ aromatic C--H stretches
of benzene, naphthalene, anthracene,
pyrene, and coronene
are much higher compared to
their gas-phase experimental results.
This is also true for methylated species
(e.g., methylated benzene or toluene,
see \S\ref{sec:B3LYP}).
Using better basis sets in conjunction with
the B3LYP method, we find that the IR intensities
still differ by a factor of $\simali$30\%
compared to the experimental results.
In contrast, Pavlyuchko et al.\ (2012)
reported that the IR intensities calculated
for benzene and toluene
at the level {\rm MP2/6-311G(3df,3pd)}
would match the experimental results very well.

Ideally, in order to compute $\Aaro$ and $\Aali$
as accurately as possible, one should study
the candidate UIE carriers at the most pertinent
levels [e.g., MP2 in conjunction with
{\rm 6-311++G$^{\ast\ast}$},
{\rm 6-311+G(3df,3pd)}, or
{\rm 6-311++G(3df,3pd)}].
However, the huge computational demand
required by these techniques often makes
it impractical to compute $\Aaro$ and $\Aali$,
particularly for large molecules.
In this work, based on methylated aromatic
hydrocarbon molecules
(with the methyl group taken to represent
the aliphatic component of the UIE carriers),
we present in \S\ref{sec:scaling}
an intensity scaling approach which,
by scaling the intensities computed
at an inexpensive level
(e.g., {\rm B3LYP/6-31G$^{\ast}$})
we are able to obtain intensities
as accurate as that computed at
far more expensive levels
(e.g., MP2/{\rm 6-311+G(3df,3pd)}).
We calculate in \S\ref{sec:astro}
the model emission spectra of PAHs
containing various numbers of methyl sidegroups,
excited by starlight of different spectral shapes
and intensities.
We derive $\left(\Iali/\Iaro\right)_{\rm mod}$,
the ratio of the model intensity of
the 3.4$\mum$ feature
to that of the 3.3$\mum$ feature.
We explore the variation of
$\left(\Iali/\Iaro\right)_{\rm mod}$
with the spectral shape and intensity
of the exciting starlight.
We summarize the principal results
in \S\ref{sec:summary}.

\section{B3LYP IR Intensities
         for C--H Stretching Modes
         \label{sec:B3LYP}}
To derive the intrinsic oscillator strengths
of the 3.3$\mum$ aromatic C--C stretch ($\Aaro$)
and the 3.4$\mum$ aliphatic C--H stretch ($\Aali$),
we have employed density functional theory
and second-order perturbation theory
to compute the IR vibrational spectra of
seven PAH species
(benzene C$_6$H$_6$,
naphthalene C$_{10}$H$_8$,
anthracene C$_{14}$H$_{10}$,
phenanthrene C$_{14}$H$_{10}$,
pyrene C$_{16}$H$_{10}$,
perylene C$_{20}$H$_{12}$,
and coronene C$_{24}$H$_{12}$),
as well as all of their methyl derivatives
(see Yang et al.\ 2013).
All of the molecules have been studied
in all conformations
at the {\rm B3LYP/6-31G$^{\ast}$} level.
The calculations always show three methyl
C--H stretches for all the methyl derivatives
of all the molecules, and we always describe
these three bands as $\nuMea$, $\nuMeb$, and $\nuMec$.

For benzene, the gas-phase experimental spectrum of
the {\it National Institute of Standards and Technology}
(NIST)\footnote{%
   The intensities for benzene are taken from
   the 3-term Blackman-Harris entries
   with a resolution of 0.125$\cm^{-1}$.
   }
gives an absorption intensity of $\simali$54.4$\km\mol^{-1}$
for the aromatic C--H stretches,
in close agreement with
the intensity of $\simali$55$\km\mol^{-1}$
computed by Pavlyuchko et al.\ (2012)
at the {\rm MP2/6-311G(3df,3pd)} level,\footnote{
   Bertie \& Keefe (1994) gave a significantly
   higher value of
   $A_{\rm aro}(\nu_{12})\approx 73\pm{9}\km\mol^{-1}$
   based on their integration over
   the range of 3175--2925$\cm^{-1}$.
   Note that this region contains some intensity
   from the (weak) combination bands.
   }
but much lower than the computed intensity of
$\simali$104$\km\mol^{-1}$
derived at the {\rm B3LYP/6-31G$^{\ast}$} level.
The gas-phase intensity measurements
of the aromatic C--H stretches
have been reported for naphthalene
($\simali$96$\km\mol^{-1}$;
Can\'{e} et al.\ 1996),
anthracene ($\simali$161$\km\mol^{-1}$;
Can\'{e} et al.\ 1997),
pyrene ($\simali$122$\km\mol^{-1}$;
Joblin et al.\ 1994),
and coronene ($\simali$161$\km\mol^{-1}$;
Joblin et al.\ 1994).
To our knowledge, no gas phase IR intensities
have been published for phenanthrene and perylene,
although the IR absorption spectra of 
various matrix-isolated PAH species 
including phenanthrene and perylene
have been obtained 
(e.g., see Hudgins \& Allamandola 1995a,b, 1997; 
Hudgins \& Sandford 1998a,b;
Szczepanski \& Vala 1993a,b). 
Similar to benzene, the experimental intensities
are much lower than our calculated results
for the aromatic C--H stretches
at the {\rm B3LYP/6-31G$^{\ast}$} level
which are respectively
$\simali$139, 178, 188 and 257$\km\mol^{-1}$
for naphthalene, anthracene, pyrene and coronene,
exceeding their experimental values
by $\simali$45\%, 11\%, 54\% and 60\%, respectively.
%



For toluene,
we digitize the NIST experimental spectra
and integrate over the  range of
3000--3200$\cm^{-1}$
to obtain the intensity of
the aromatic C--H stretch ($A_{\rm aro}$).
Similarly, we integrate over the  range of
2800--3000$\cm^{-1}$
to obtain the intensity of
the aliphatic C--H stretch ($A_{\rm ali}$).
%
The relative intensity of the methyl (aliphatic) signal
to that of the aromatic band is $\Arel \approx 0.79$.
A similar analysis of the experimental
spectrum of Wilmshurst \& Bernstein (1957)
results in $\Arel\approx0.71$.\footnote{%
  Note that $A_{\rm aro}$ ($A_{\rm ali}$)
  is the strength of all the aromatic (aliphatic)
  C--H stretches while $\Aaro$ ($\Aali$) is
  the strength of the aromatic (aliphatic) stretch
  per C--H bond. For toluene,  $A_{\rm aro} = 5\Aaro$
  and $A_{\rm ali} = 3\Aali$ and therefore
  we have $\Aratio = \left(5/3\right)\,\Arel$.
  }
Our integration of the NIST spectrum of toluene
gives a total intensity
of $\simali$97.2$\km\mol^{-1}$
for all the C--H stretches (both methyl and aromatic)
and is in excellent agreement with the value
of $\simali$95$\km\mol^{-1}$ calculated by
Pavlyuchko et al.\ (2012)
and by Galabov et al.\ (1992)
at the {\rm MP2/6-311G(3df,3pd)} level.
According to our ratio of the measured intensities
for the methyl to aromatic regions
($\Arel \approx 0.79$),
this overall intensity corresponds to intensities
of $\simali$42.9$\km\mol^{-1}$ for the methyl bands
and of $\simali$54.3$\km\mol^{-1}$ for the aromatic bands.
The intensities computed
at the {\rm B3LYP/6-31G$^{\ast}$} level
for toluene are $\simali$165.3$\km\mol^{-1}$
for the entire region
and $\simali$70.4 and $\simali$94.9$\km\mol^{-1}$
for the methyl and aromatic sections, respectively.
Again, we see that the computed intensities
are much higher than the experimental values
from the gas phase measurements.

In the absence of absolute intensity experimental
data for naphthalene, anthracene, phenanthrene,
perylene, pyrene and coronene,
we are unfortunately not able to compare
the experimental intensities of the C--H stretches
of these molecules with that computed at
the {\rm B3LYP/6-31G$^{\ast}$} level.
%
%

\section{Scaling Approaches for
         the Computed Total Intensities
         of C--H Stretching Modes
         \label{sec:scaling}
         }
As we have seen in \S\ref{sec:B3LYP},
the IR intensities calculated
at the {\rm B3LYP/6-31G$^{\ast}$} level
are much higher
compared to the experimental results.
Using better basis sets in conjunction with
the B3LYP method, we found that the IR intensities
still differ by a factor of $\simali$30\%
compared to the experiment results.
Pavlyuchko et al.\ (2012) reported that
the IR intensities calculated for benzene and toluene
at the level {\rm MP2/6-311G(3df,3pd)} would match
the experimental results very well.
We have tried to reproduce their data for benzene
and toluene by performing both MP2(fc)
and MP2(full) computations
with the 6-311G(3df,3pd) basis set.\footnote{%
  The MP2 computations are performed
  either with the full active space of
  all core and valence electrons
  considered in the correlation energy computation,
  denoted MP2(full),
  or with the frozen core approximation
  and the consideration of just the valence electrons
  in the correlation treatment, denoted MP2(fc).
   With MP2/6-311G(3df,3pd),
   Pavlyuchko et al.\ (2012) calculated 
   the C--H stretch intensities of benzene 
   and toluene to be $\simali$53$\km\mol^{-1}$
   and $\simali$98$\km\mol^{-1}$, respectively.
   We have tried both MP2(fc)/6-311G(3df,3pd) 
   and MP2(full)/6-311G(3df,3pd).
   With MP2(fc)/6-311G(3df,3pd),
   we obtained $\simali$53.8$\km\mol^{-1}$
   and $\simali$97.1$\km\mol^{-1}$
   for benzene and toluene, respectively,
   while with MP2(full)/6-311G(3df,3pd)
   these  intensities become 
   $\simali$52.4$\km\mol^{-1}$
   and $\simali$94.7$\km\mol^{-1}$.
   Although the MP2(fc) results closely 
   match that of Pavlyuchko et al.\ (2012),
   the MP2(full) results are closer to 
   the experimental results
   ($\simali$55$\km\mol^{-1}$ for benzene
    and $\simali$95$\km\mol^{-1}$ for toluene).
   Since MP2(full) considers all the core 
   and valence electrons and thus should be more 
   accurate than MP2(fc), we therefore calculate all 
   other vibrational spectra with MP2(full)
   in conjunction with the standard basis set 
   6-31G$^{\ast}$ and the extended basis sets 
   6-311+G$^{\ast\ast}$ and 6-311+G(3df,3pd) 
   for benzene, naphthalene and their mono-methyl 
   derivatives as test cases.
   }
%

While the MP2(full)/6-311+G(3df,3pd) level data
reproduce the measured IR intensities reasonably well,
such calculations are far too expensive especially
for large molecules. The MP2(full) computations of
the naphthalene systems with the large basis sets
including the (3df,3pd) polarization functions
each requires several days of computer time on
eight processors. Considering that the absolute values
computed at all of the MP2 levels are better than
the respective values computed at the B3LYP levels,
one would be inclined to explore scaling approaches
of the MP2 data computed with modest basis sets.
However, we will show below that scaling approaches
that are based on the B3LYP data
can be just as successful
in spite of the fact that the absolute numbers
computed at the {\rm B3LYP/6-31G$^{\ast}$} level
differ much more from experiment than
do the {\rm MP2/6-31G$^{\ast}$} data.

Before we proceed, it is useful to clarify
the meaning of scaling approaches.
In the most typical approach to scaling,
it is attempted to reproduce a set
of experimental data with a set of data
obtained at a level $\Li$ such that
$p({\rm exp}) \approx f \cdot p(\Li)$,
that is, one scaling factor $f$ is applied
to all values in the data set and
this scale factor depends on the level, $f = f(\Li)$.
This kind of scaling is commonly employed for
vibrational frequencies.
For intensities, however,
we will see that approaches of
the type $p({\rm exp}) \approx f \cdot p(\Li)+C(\Li)$
are more successful, that is,
there will be a non-zero offset.

\begin{figure*}
\centerline
{
\includegraphics[scale=0.6,clip]{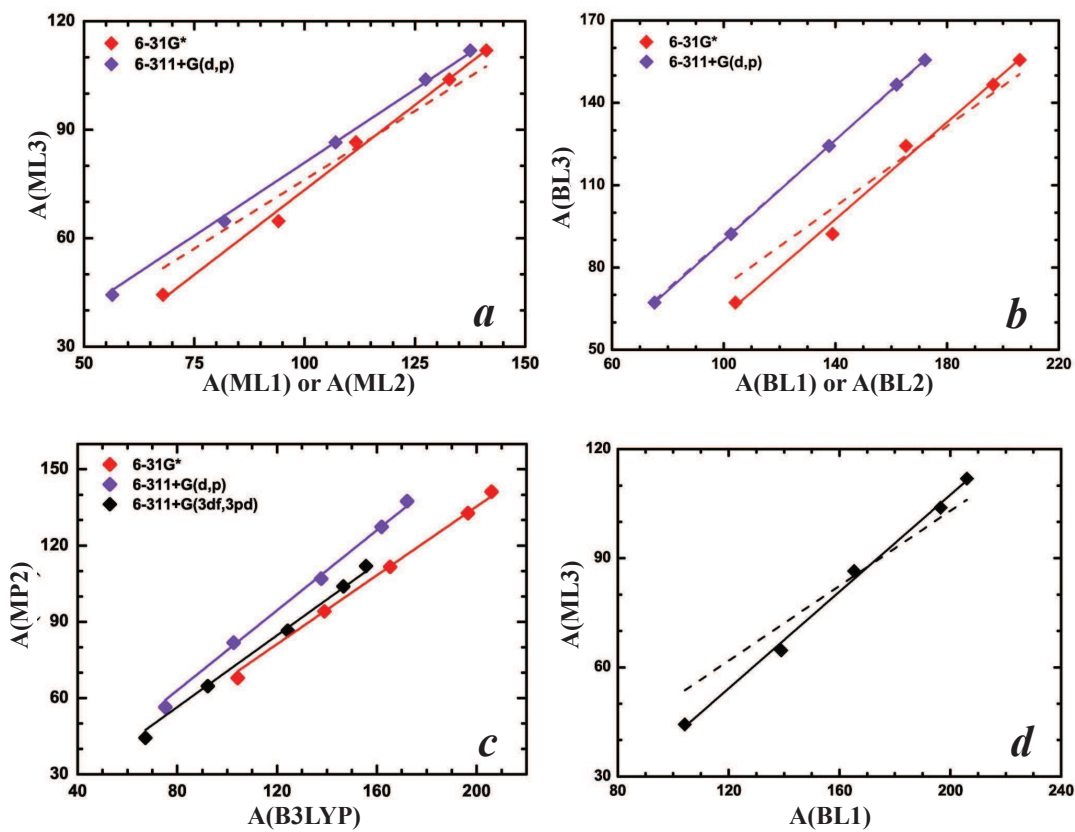}
}
\caption{\footnotesize
         \label{fig:Int_LevelDep}
         Level dependency of the total C--H stretch
         intensities (methyl plus aromatic)
         for benzene and naphthalene and for their methyl
         derivatives toluene and methylnaphthalene.
         Top left (a): Intensities calculated at MP2
         with small basis sets
         [i.e., 6-31G$^{\ast}$ (i.e., $ML1$),
          6-311+G(d,p) (i.e., $ML2$)]
         vs. that with a large basis set
         [6-311+G(3df,3pd) (i.e., $ML3$)].
         Dashed red line plots eq.\,\ref{eq:MP2_LS_SS_a},
         solid red line plots eq.\,\ref{eq:MP2_LS_SS_b},
         and solid blue line plots eq.\,\ref{eq:MP2_LS_MS}.
         Top right (b): Same as (a) but at B3LYP.
         Dashed red line plots eq.\,\ref{eq:B3LYP_LS_MS_a},
         solid red line plots eq.\,\ref{eq:B3LYP_LS_MS_b},
         dashed blue line plots eq.\,\ref{eq:B3LYP_LS_SS_a},
         and solid blue line plots eq.\,\ref{eq:B3LYP_LS_SS_b}.
         Bottom left (c): Intensities calculated
         at B3LYP vs. MP2 with the same basis set.
         Solid red line plots eq.\,\ref{eq:B3LYP_MP2_SS},
         solid blue line plots eq.\,\ref{eq:B3LYP_MP2_MS},
         and solid black line plots eq.\,\ref{eq:B3LYP_MP2_LS}.
         Bottom right (d): Intensities calculated
         at B3LYP/6-31G$^{\ast}$ (i.e., $BL1$)
         vs. MP2/6-311+G(3df,3pd) (i.e., $ML3$).
         Dashed black line plots eq.\,\ref{eq:B3LYP_MP2_a},
         and solid black line plots eq.\,\ref{eq:B3LYP_MP2_b}
         }
\end{figure*}

Let $ML1$, $ML2$ and $ML3$ respectively
represent the {\rm MP2(full)} computations
with the {\rm 6-31G$^{\ast}$}, 6-311+G(d,p),
and 6-311+G(3df,3pd) basis sets.
%
Let $BL1$, $BL2$ and $BL3$ respectively
represent the B3LYP computations
with the {\rm 6-31G$^{\ast}$}, 6-311+G(d,p),
and 6-311+G(3df,3pd) basis sets.
%
As can be seen from Figure~\ref{fig:Int_LevelDep} (top left),
the total intensities ($A$) computed at the MP2 level
but with different basis sets
[i.e., $A(ML1)$, $A(ML2)$, and $A(ML3)$]
are linearly related:
\begin{subequations}\label{eq:MP2_LS_SSMS}
\begin{align}
A(ML3)&\approx0.7615\,A(ML1)~~,~~(r^{2}\approx0.9575)
\label{eq:MP2_LS_SS_a} \\
A(ML3)&\approx0.9382\,A(ML1)-20.4880~~,~~(r^{2}\approx0.9949)
\label{eq:MP2_LS_SS_b} \\
A(ML3)&\approx0.8089\,A(ML2)~~,~~(r^{2}\approx0.9984)
\label{eq:MP2_LS_MS}
\end{align}
\end{subequations}
where $r^{2}$ is the linear-correlation coefficient.
While eq.\,\ref{eq:MP2_LS_MS} describes
an excellent linear correlation
between the intensities
computed with the $ML3$ method 
and that with the $ML2$ method 
without any need for an offset,
the analogous eq.\,\ref{eq:MP2_LS_SS_a}
is less successful and an excellent linear
correlation between $A(ML3)$ and $A(ML1)$
only is achieved when a non-zero offset
is allowed in eq.\,\ref{eq:MP2_LS_SS_b}.
The analogous relations also hold at the B3LYP level
(eq.\,\ref{eq:B3LYP_LS_SS})
and they are shown in
Figure~\ref{fig:Int_LevelDep} (top right),
where $A(BL1)$, $A(BL2)$, and $A(BL3)$ are
respectively the intensities computed at
the $BL1$, $BL2$ and $BL3$ levels.
%
%
%
%
\begin{subequations}\label{eq:B3LYP_LS_SS}
\begin{align}
A(BL3)&\approx0.7306\,A(BL1)~,~~(r^{2}\approx0.9610)
\label{eq:B3LYP_LS_MS_a} \\
A(BL3)&\approx0.8838\,A(BL1)-26.1670~,~~(r^{2}\approx0.9924)
\label{eq:B3LYP_LS_MS_b}\\
A(BL3)&\approx0.8089\,A(BL2)~,~~(r^{2}\approx0.9984)
\label{eq:B3LYP_LS_SS_a} \\
A(BL3)&\approx0.8395\,A(BL2)-3.3861~,~~(r^{2}\approx0.9998)
\label{eq:B3LYP_LS_SS_b}
\end{align}
\end{subequations}
Also shown in Figure~\ref{fig:Int_LevelDep} (bottom left)
are the nearly linear relations between the IR intensities
computed at the B3LYP and MP2(full) levels
with a common basis set.
The data are very well described by linear regression
and there is no need for a non-zero offset
in any of the following equations
(see eqs.\,\ref{eq:B3LYP_MP2_SS},
\ref{eq:B3LYP_MP2_MS}, and \ref{eq:B3LYP_MP2_LS}).
It is remarkable that these slopes are rather similar
for the various basis sets.
\begin{subequations} \label{eq:B3LYP_MP2_scale}
\begin{align}
A(ML1)&\approx0.6769\,A(BL1)~,~~(r^{2}\approx0.9971)
\label{eq:B3LYP_MP2_SS} \\
A(ML2)&\approx0.7877\,A(BL2)~,~~(r^{2}\approx0.9966)
\label{eq:B3LYP_MP2_MS} \\
A(ML3)&\approx0.7056\,A(BL3)~,~~(r^{2}\approx0.9949)
\label{eq:B3LYP_MP2_LS}
\end{align}
\end{subequations}

In light of these linear correlations,
it is clear that there must be a strong
linear correlation between the lowest DFT level,
our standard level {\rm B3LYP/6-31G${^{\ast}}$}
(i.e., $BL1$), and the best MP2 level,
the level MP2(full)/6-311+G(3df,3pd)
(i.e., $ML3$).
Eqs.\,\ref{eq:MP2_LS_SS_a}
and \ref{eq:B3LYP_MP2_SS}
suggest a correlation coefficient of
$\approx 0.7615\times0.6769\approx 0.5154$
and the actual correlation coefficient of
eq.\,\ref{eq:B3LYP_MP2_a} is $\simali$0.5152
and it is essentially the same
(see Figure~\ref{fig:Int_LevelDep}, bottom right).
Considering the need for non-zero offset in
eq.\,\ref{eq:MP2_LS_SS_b},
we also explore eq.\,\ref{eq:B3LYP_MP2_b}
and achieve an excellent linear correlation:
\begin{subequations}\label{eq:B3LYP_MP2}
\begin{align}
A(ML3)&\approx0.5152\,A(BL1)~,~~(r^{2}\approx0.9428)
\label{eq:B3LYP_MP2_a}  \\
A(ML3)&\approx0.6655\,A(BL1)-25.6770~,~~(r^{2}\approx0.9964)
\label{eq:B3LYP_MP2_b}
\end{align}
\end{subequations}
This tells that, by applying this scaling relation
(eq.\,\ref{eq:B3LYP_MP2_b}), we just need to
perform computations at an inexpensive
level (e.g., {\rm B3LYP/6-31G$^{\ast}$})
and we are still able to obtain intensities
as accurate as that computed at
far more advanced levels
[e.g., MP2/{\rm 6-311+G(3df,3pd)}].

\section{Astrophysical Implications
            \label{sec:astro}}
As shown in Yang et al.\ (2013),
the aromatic C--H stretch band strength
does not vary significantly for different molecules.
It has an average value (per aromatic C--H bond) of
$\langle \Aaro\rangle \approx 14.03\km\mol^{-1}$,
with a standard deviation of
$\sigma(\Aaro)\approx 0.89\km\mol^{-1}$.
On the other hand, the aliphatic C--H stretch
band strength is more dependent on the nature
of the molecule and also on the specific isomer.
The average band strength (per aliphatic C--H bond)
is $\langle \Aali\rangle \approx 23.68\km\mol^{-1}$,
and the standard deviation is
$\sigma(\Aali)\approx 2.48\km\mol^{-1}$.
All of these values are calculated
for neutral PAHs
at the {\rm B3LYP/6-311+G$^{\ast\ast}$}
(i.e., $BL2$) level.
As discussed in \S\ref{sec:scaling}, these values
need to be scaled.
By taking {\rm MP2(full)/6-311+G(3df,3pd)}
(i.e., $ML3$) to be the level
which gives the most reliable band strength,
the intensities need to be scaled with two formulae:
eqs.\,\ref{eq:MP2_LS_MS} and \ref{eq:B3LYP_MP2_MS}.
Thus, we derive for neutral PAHs
$\langle\Aaro\rangle \approx 14.03
\times 0.7877 \times 0.8089
\approx 8.94\km\mol^{-1}$
(i.e., $\simali$$1.49\times10^{-18}\cm$
per C--H bond),
$\langle \Aali\rangle \approx 23.68
\times 0.7877 \times 0.8089
\approx 15.09\km\mol^{-1}$
(i.e., $\simali$$2.50\times10^{-18}\cm$
per C--H bond), and
$\langle\Aali\rangle/\langle\Aaro\rangle \approx 1.69$.
Similarly, we obtain for PAH cations
$\langle\Aaro\rangle
\approx 0.92\km\mol^{-1}$,
$\langle \Aali\rangle
\approx 3.20\km\mol^{-1}$, and
$\langle\Aali\rangle/
\langle\Aaro\rangle \approx 3.48$.
We note that, although these results
were derived from the mono-methyl 
derivatives of small PAH molecules,
it has been shown in Yang et al.\ (2016b)
that the $\Aratio$ ratios determined 
from the PAH molecules attached with 
a wide range of sidegroups
(including ethyl, propyl, and butyl)
as well as dimethyl-substituted pyrene
are close to that of mono-methyl PAHs. 

In addition to the 3.4$\mum$ C--H stretch,
PAHs with aliphatic sidegroups also have two
aliphatic C--H deformation bands
at 6.85$\mum$ and 7.25$\mum$.
Yang et al.\ (2016a) have derived
$\Adfa$ and $\Adfb$,
the intrinsic oscillator strengths
of the 6.85 and 7.25$\mum$
aliphatic C--H deformation bands
for both neutral and ionized
methyl-substituted PAHs.
They obtained lower limits of
$\Adfa/\Acc\approx5.0$
and $\Adfb/\Acc\approx0.5$ for neutrals,
$\Adfa/\Acc\approx0.5$
and $\Adfb/\Acc\approx0.25$ for cations,
where $\Acc$ is the intrinsic oscillator
strength of the 6.2$\mum$ aromatic
C--C stretch.

With $\Aali/\Aaro$, $\Adfa/\Acc$ and $\Adfb/\Acc$
derived for both neutral and ionized PAHs,
we now calculate the emission spectra
of methyl PAHs excited by starlight
and the corresponding model band ratios
$\Iali/\Iaro$.
Lets consider a PAH molecule containing
$N_{\rm C,aro}$ aromatic C atoms and
$N_{\rm C,ali}$ aliphatic C atoms
(i.e., $N_{\rm C,ali}$ methyl sidegroups).
We approximate
their absorption cross sections by
adding three Drude functions to that
of PAHs of $\NCaro$ aromatic C atoms,
with these Drude functions respectively
representing the 3.4$\mum$ aliphatic
C--H stretch, and the 6.85 and 7.25$\mum$
aliphatic C--H deformations:
\begin{eqnarray}
C_{\rm abs}(\NC,\lambda) & = & \CabsPAH(\NCaro,\lambda)\\
& + & \NCali \frac{2}{\pi}
    \frac{\gamma_{3.4} \lambda_{3.4} \sigma_{\rm int,3.3}
     \left(A_{3.4}/A_{3.3}\right)}
     {(\lambda/\lambda_{3.4}-\lambda_{3.4}/\lambda)^2
      +\gamma_{3.4}^2}\\
&+& \NCali \frac{2}{\pi}
     \frac{\gamma_{6.85} \lambda_{6.85}
     \sigma_{\rm int,6.2} \left(\Adfa/\Acc\right)}
     {(\lambda/\lambda_{6.85}-\lambda_{6.85}/\lambda)^2
      +\gamma_{6.85}^2}\\
&+& \NCali \frac{2}{\pi}
    \frac{\gamma_{7.25} \lambda_{7.25}
    \sigma_{\rm int,6.2} \left(\Adfb/\Acc\right)}
     {(\lambda/\lambda_{7.25}-\lambda_{7.25}/\lambda)^2
      +\gamma_{7.25}^2} ,
\end{eqnarray}
where $\NC=\NCaro+\NCali$; $\lambda_{3.4}=3.4\mum$,
$\lambda_{6.85}=6.85\mum$, and $\lambda_{7.25}=7.25\mum$
are respectively the peak wavelengths
of the 3.4, 6.85 and 7.25$\mum$ features;
$\gamma_{3.4}\lambda_{3.4}=0.03\mum$,
$\gamma_{6.85}\lambda_{6.85}=0.2\mum$,
and $\gamma_{7.25}\lambda_{7.25}=0.2\mum$
are respectively the mean FWHMs of the
astronomical 3.4, 6.85 and 7.25$\mum$ features
(Yang et al.\ 2003, 2016a),\footnote{%
   As defined by Draine \& Li (2007),
  $\gamma_{3.4}$, $\gamma_{6.85}$, and
  $\gamma_{7.25}$ are dimentionless parameters.
  }
and $\sigma_{{\rm int},3.3}$ and $\sigma_{{\rm int},6.2}$
are respectively the integrated strengths per (aromatic)
C atom of the 3.3$\mum$ aromatic C--H stretch
and 6.2$\mum$ aromatic C--C stretch
(see Draine \& Li 2007).

\begin{figure*}
\begin{center}
\includegraphics[scale=0.7,clip]{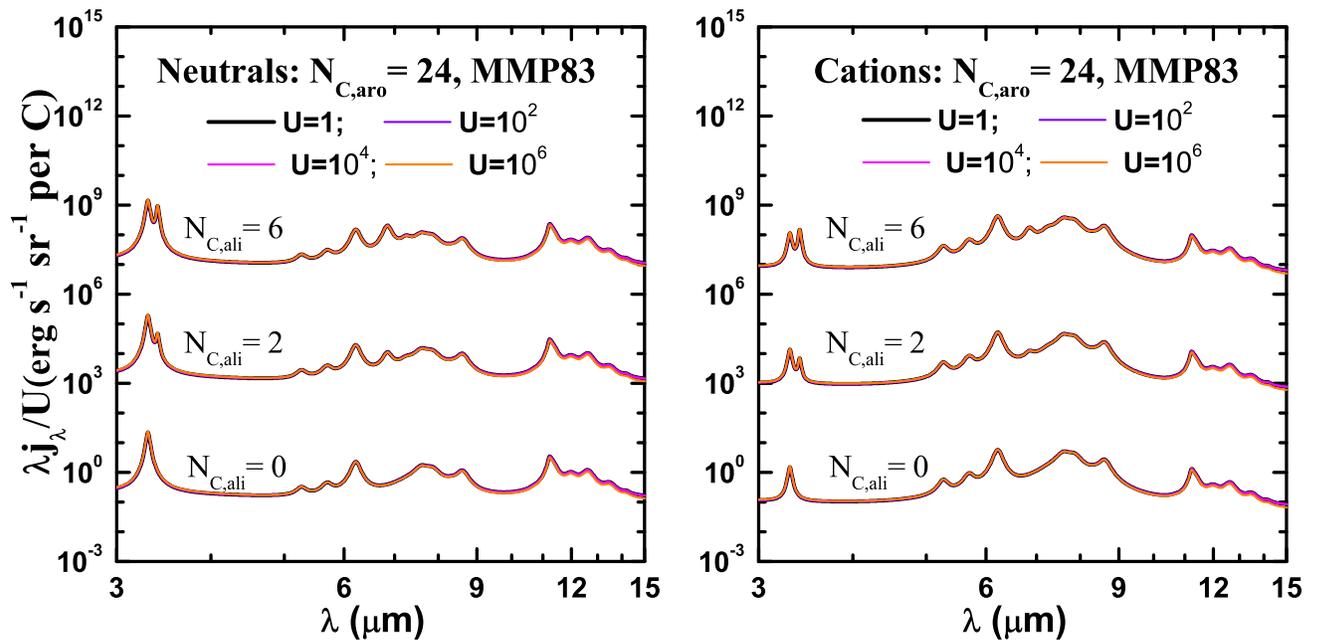}
\caption{\footnotesize
         \label{fig:modspec_U}
         IR emission spectra of neutral (left panel)
         and ionized (right panel) methyl PAHs
         of $\NCali=0, 2, 6$ aliphatic C atoms
         and $\NCaro=24$ aromatic C atoms
         illuminated by the MMP83 ISRF of
         various intensities ($U=1$: black lines;
         $U=100$: purple lines; $U=10^4$: magenta lines;
         and  $U=10^6$: red lines).
         The 3.4 and 6.85$\mum$ aliphatic C--H features
         are clearly seen in the spectra of methyl PAHs
         with $\NCali=2, 6$, while the 7.25$\mum$ aliphatic
         C--H feature is less prominent.
         For clarity, their spectra
         are vertically shifted.
         }
\end{center}
\end{figure*}

Due to their small size (and therefore small heat
capacity), PAHs are heated sporadically by single
starlight photons. Unless exposed to an extremely
intense radiation field, PAHs will undergo strong
temperature fluctuations and will not attain an
equilibrium temperature (see Li 2004).
We take the ``thermal-discrete'' technique
developed by Draine \& Li (2001) to calculate
the temperature probability distribution functions
and the resulting emission spectra of methyl PAHs.
Let $dP$ be the probability that the temperature
of the molecule will be in $[T,T+dT]$.
The emissivity of this molecule
(of $N_{\rm C}$ C atoms) becomes
\begin{equation}
j_\lambda(\NC) = \int C_{\rm abs}(\NC,\lambda)\,
            4\pi B_\lambda(T)\,\frac{dP}{dT}\,dT  ~,
\end{equation}
where $B_\lambda\left(T\right)$ is
the Planck function at wavelength $\lambda$
and temperature $T$.
As shown in Figures~6, 7 of Draine \& Li (2007),
the 3.3$\mum$ interstellar UIE emitters  are
in the size range of $\NC$\,$\simali$20--30 C atoms.
For illustrative purpose, we therefore consider
$N_{\rm C,aro}=24$ (like coronene).
For a coronene-like molecule, up to 12 methyl
sidegroups can be attached to it.
We thus consider methyl PAHs of
$\NCali=0, 1, 2, ..., 12$ aliphatic C atoms.
For all molecules, we fix $N_{\rm C,aro}=24$.
In Figure~\ref{fig:modspec_U} we show
the IR emission spectra of both neutral
and ionized methyl PAHs of $\NCali=0, 2, 6$
illuminated by the solar neighbourhood
interstellar radiation field (ISRF) of
Mathis, Mezger \& Panagia (1983; MMP83).
Figure~\ref{fig:modspec_U} shows that,
the 3.4 and 6.85$\mum$ features
are clearly visible in the IR emission spectra
for $\NCali=2$, while the 7.25$\mum$ feature
remains hardly noticeable even for $\NCali=6$.
This is because the intrinsic strength of
the 7.25$\mum$ feature is weaker than that of
the 6.85$\mum$ feature by a factor
of $\simali$8 for neutral methyl PAHs
and by a factor of $\simali$3 for their cations
(Yang et al.\ 2016a). In the following discussions,
we will focus on the 3.3 and 3.4$\mum$ features
since the molecules considered here are too small
to be the dominant UIE emitters
at $\simali$6--8$\mum$
(see Figures~6, 7 of Draine \& Li 2007).

We have also explored the effects of starlight intensities
on the IR emission spectra of methyl PAHs by increasing
the MMP ISRF by a factor of $U$.
As shown in Figure~\ref{fig:modspec_U},
the resulting IR emission spectra
for $U=1, 100, 10^4, 10^6$,
after scaled by $U$,
are essentially identical.
This is not unexpected.
The single-photon heating nature of
these molecules assures that their IR
emission spectra (scaled by the starlight intensity)
to remain the same for different starlight intensities.
Single-photon heating implies that the shape of the
high-$T$ end of the temperature probability distribution
function $dP/dT$ for a methyl PAH 
is the same for different levels
of starlight intensity, and what only matters is the mean
photon energy (which determines to what peak temperature
a molecule will reach, upon absorption of such a photon;
see Draine \& Li 2001, Li 2004).

\begin{figure*}
\begin{center}
\includegraphics[scale=0.7,clip]{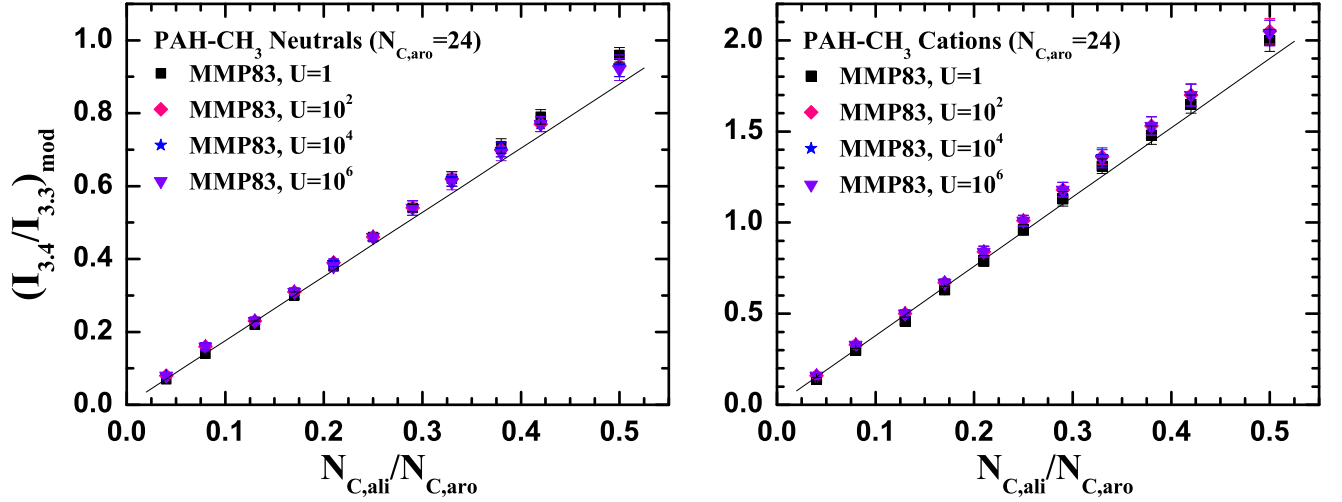}
\caption{\footnotesize
         \label{fig:Iratio_U}
              Model-calculated intensity ratios $\Iratiomod$
              as a function of the aliphatic fraction $\NCali/\NCaro$
              for neutral methyl PAHs of $\NCaro=24$ (left panel)
              and their cations (right panel).
              The molecules and their cations are illuminated by
              the MMP83 ISRF with the starlight intensity enhanced
              by a factor of $U$
              ($U=1$: black squares; $U=100$: red diamonds;
               $U=10^4$: blue stars; $U=10^6$: purple triangles).
              The solid black line plots $\Iratiomod =1.76\times
               \left(\NCali/\NCaro\right)$
               for neutrals and $\Iratiomod =3.80\times
               \left(\NCali/\NCaro\right)$
               for cations.
               }
\end{center}
\end{figure*}

For a given $\NCali$, we derive $\Iratiomod$,
the model intensity ratio of the 3.4$\mum$ band
to the 3.3$\mum$ band, from
\begin{equation}\label{eq:Iratiomod}
\left(\frac{I_{3.4}}{I_{3.3}}\right)_{\rm mod}
= \frac{\int_{3.4}\Delta j_\lambda(\NC)\,d\lambda}
{\int_{3.3}\Delta j_\lambda(\NC)\,d\lambda} ~~,
\end{equation}
where $\int_{3.3}\Delta j_\lambda(\NC)\,d\lambda$
and $\int_{3.4}\Delta j_\lambda(\NC)\,d\lambda$
are respectively the feature-integrated excess emission
of the 3.3 and 3.4$\mum$ features 
of the methyl PAH molecule. 
In Figure~\ref{fig:Iratio_U} we show
the model intensity ratios $\Iratiomod$
as a function of $\NCali/\NCaro$
for neutral and ionized methyl PAHs.
It is encouraging to see
in Figure~\ref{fig:Iratio_U} that,
with $\NCali/\NCaro =0.5$,
$\Iratiomod$ reaches $\simali$0.9 for neutrals
and $\simali$2.0 for cations,
demonstrating that the unusually high
$\Iratioobs$ ratios observed in some
protoplanetary nebulae
(e.g., IRAS~04296+3429 
with $\Iratioobs\approx1.54$)
can be accounted for by a mixture of neutral
and ionized methyl PAHs, with a reasonable
fraction of C atoms in methyl sidegroups.
In Figure~\ref{fig:Iratio_U} we also compare
the model band ratios with the ratios computed
from the simple relation
$\Iratiomodp = 1.76\times\left(\NCali/\NCaro\right)$
for neutrals or 
$\Iratiomodp = 3.80\times\left(\NCali/\NCaro\right)$
for cations.
Figure~\ref{fig:Iratio_U} shows that
this simple, straightforward relation
does an excellent job in accurately
predicting $\Iratiomod$.
This is nice because in future studies
one can simply use this convenient
relation to determine the aliphatic fraction
$\NCali/\NCaro$ of the UIE carrier
from the observed band ratio $\Iratioobs$:
$\NCali/\NCaro\approx0.57\times\Iratioobs$
for neutrals and
$\NCali/\NCaro\approx0.26\times\Iratioobs$
for cations.
There is no need to compute
the temperature probability
distribution functions and
the IR emission spectra of
methyl PAHs as long as one
is only interested in the aliphatic
fraction of the UIE carrier.

\begin{figure*}
\begin{center}
\includegraphics[scale=0.5,clip]{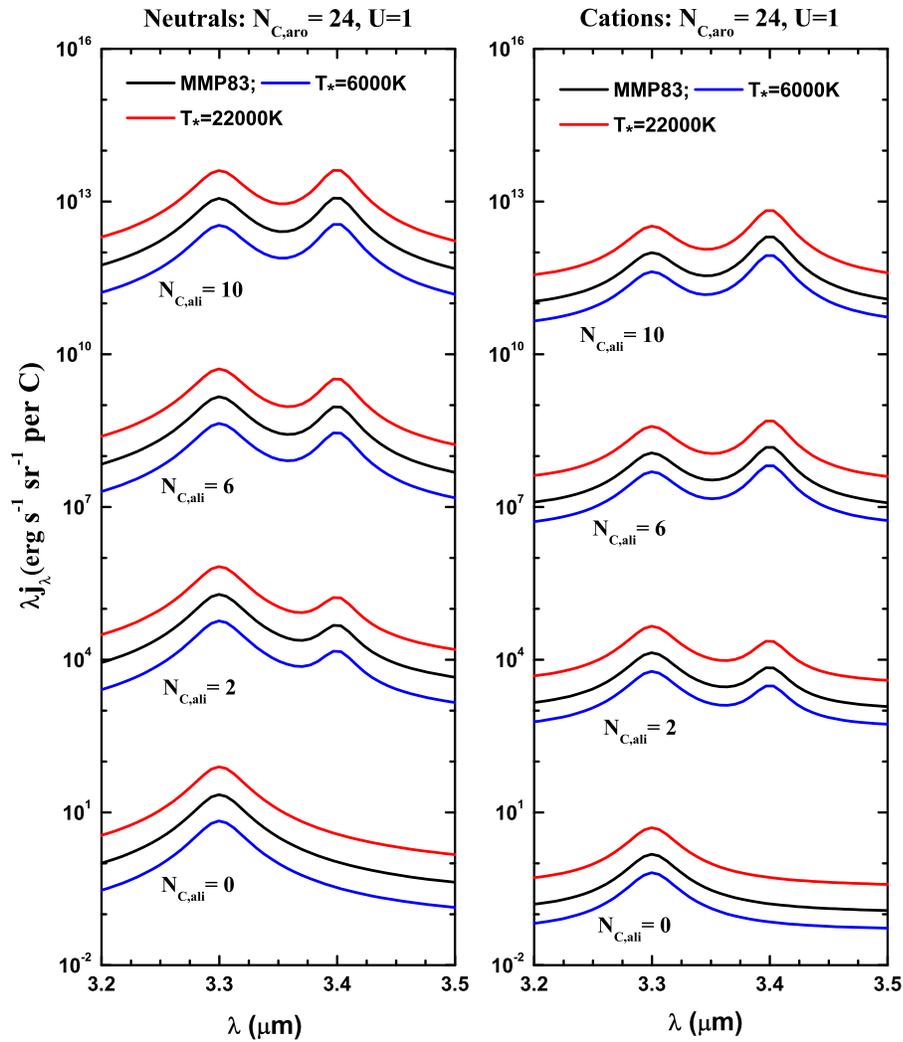}
\vspace{-4cm}
\caption{\footnotesize
         \label{fig:modspec_T}
         IR emission spectra of neutral (left panel)
         and ionized (right panel) methyl PAHs
         of $\NCali=0, 2, 6, 10$ aliphatic C atoms
         and $\NCaro=24$ aromatic C atoms
         illuminated by a solar-type star of
         $\Teff=6000\K$ (blue lines),
         a B1.5V star of $\Teff=22,000\K$ (red lines),
         and the MMP83 ISRF (black lines).
         The starlight intensities are all set
         to be $U=1$.
         For clarity, the spectra
         for methyl PAHs with $\NCali=2, 6, 10$
         are vertically shifted.
         }
\end{center}
\end{figure*}

\begin{figure*}
\begin{center}
\includegraphics[scale=0.7,clip]{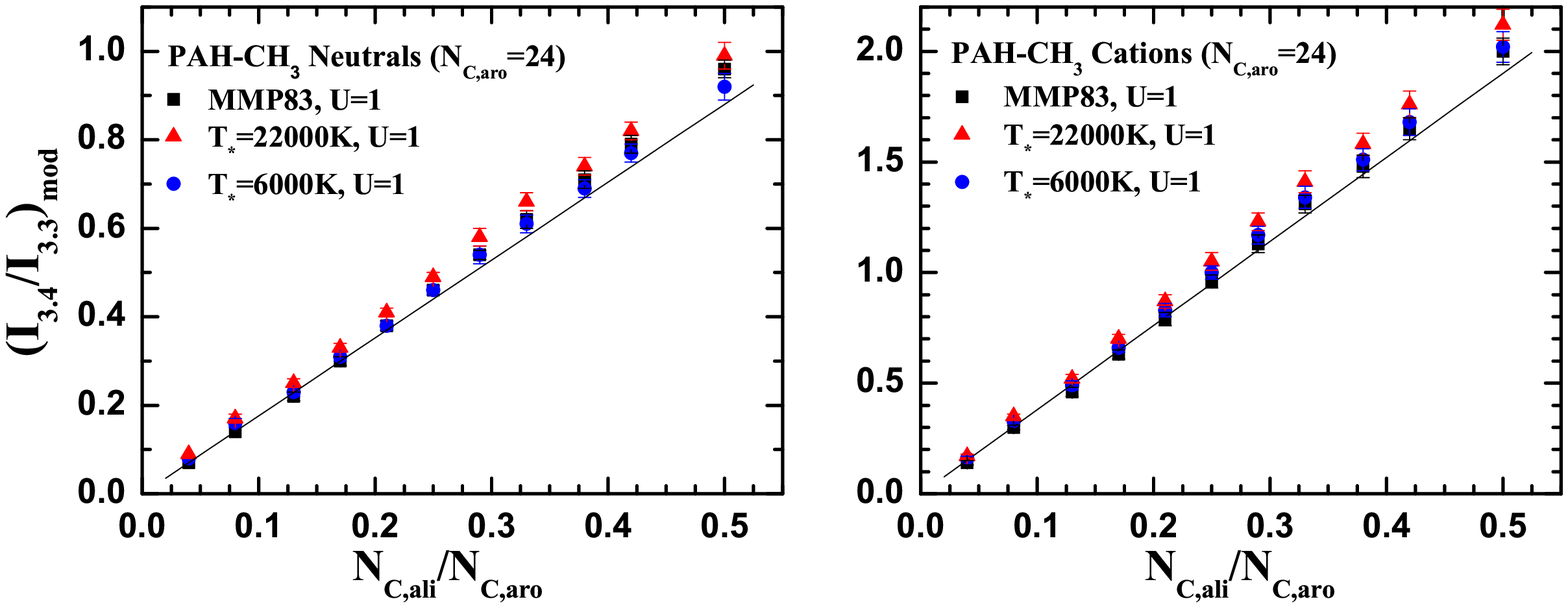}
\caption{\footnotesize
         \label{fig:Iratio_T}
              Model-calculated intensity ratios $\Iratiomod$
              as a function of the aliphatic fraction $\NCali/\NCaro$
              for neutral methyl PAHs of $\NCaro=24$ (left panel)
              and their cations (right panel).
              The molecules and their cations are illuminated by
              a solar-type star of
             $\Teff=6000\K$ (blue circles),
             a B1.5V star of $\Teff=22,000\K$ (red triangles),
             and the MMP83 ISRF (black squares).
             The starlight intensities are all set
             to be $U=1$.
             The solid black line plots $\Iratiomod =1.76\times
               \left(\NCali/\NCaro\right)$
               for neutrals and $\Iratiomod =3.80\times
               \left(\NCali/\NCaro\right)$
               for cations.
              }
\end{center}
\end{figure*}

So far, we have only considered
methyl PAHs excited by
the MMP83-type starlight.
To examine whether and how
the spectral shape of the exciting starlight
affects the model IR emission spectra
and the band ratios $\Iratiomod$,
we consider methyl PAHs
of $\NCali=0, 1, 2, ...12$ aliphatic C atoms
and $\NCali=24$ aromatic C atoms
excited by stars with an effective temperature
of $\Teff=6,000\K$ like our Sun
and by stars of $\Teff=22,000\K$
like the B1.5V star HD\,37903
which illuminates the reflection nebula NGC\,2023.
We fix the starlight intensity
in the 912$\Angstrom$--1$\mum$
wavelength range to be that of the MMP83 ISRF
(i.e., $U=1$):
\begin{equation}
\int_{1\mu {\rm m}}^{912{\rm \Angstrom}}
               4\pi J_\star(\lambda,\Teff)\,d\lambda
= \int_{1\mu {\rm m}}^{912{\rm \Angstrom}}
               4\pi J_{\rm ISRF}(\lambda)\,d\lambda ~~,
\end{equation}
where $J_\star(\lambda, \Teff)$ is the intensity
of starlight approximated by the Kurucz model
atmospheric spectrum,
and $J_{\rm ISRF}(\lambda)$
is the MMP83 ISRF starlight intensity.
As shown in Figure~\ref{fig:modspec_T},
for a given $\NCali/\NCaro$,
the $\Teff=6,000\K$ model results in
a lower emissivity level than
that of the MMP83 ISRF model.
In contrast, the $\Teff=22,000\K$ model
results in a higher emissivity level than
that of the MMP83 ISRF model.
This is because, exposed to 
a {\it softer} radiation field,
PAHs absorb individual photons 
with a {\it lower} mean energy
than that of a {\it harder} radiation field
and therefore emit less 
(because they absorb less).
Nevertheless, the emission spectral profiles
are very similar to each other.
This is also illustrated in Figure~\ref{fig:Iratio_T}
which shows that the model band ratios
$\Iratiomod$ differ very little for methyl PAHs
excited by starlight of different spectral shapes.

\begin{figure*}
\begin{center}
\includegraphics[scale=0.5,clip]{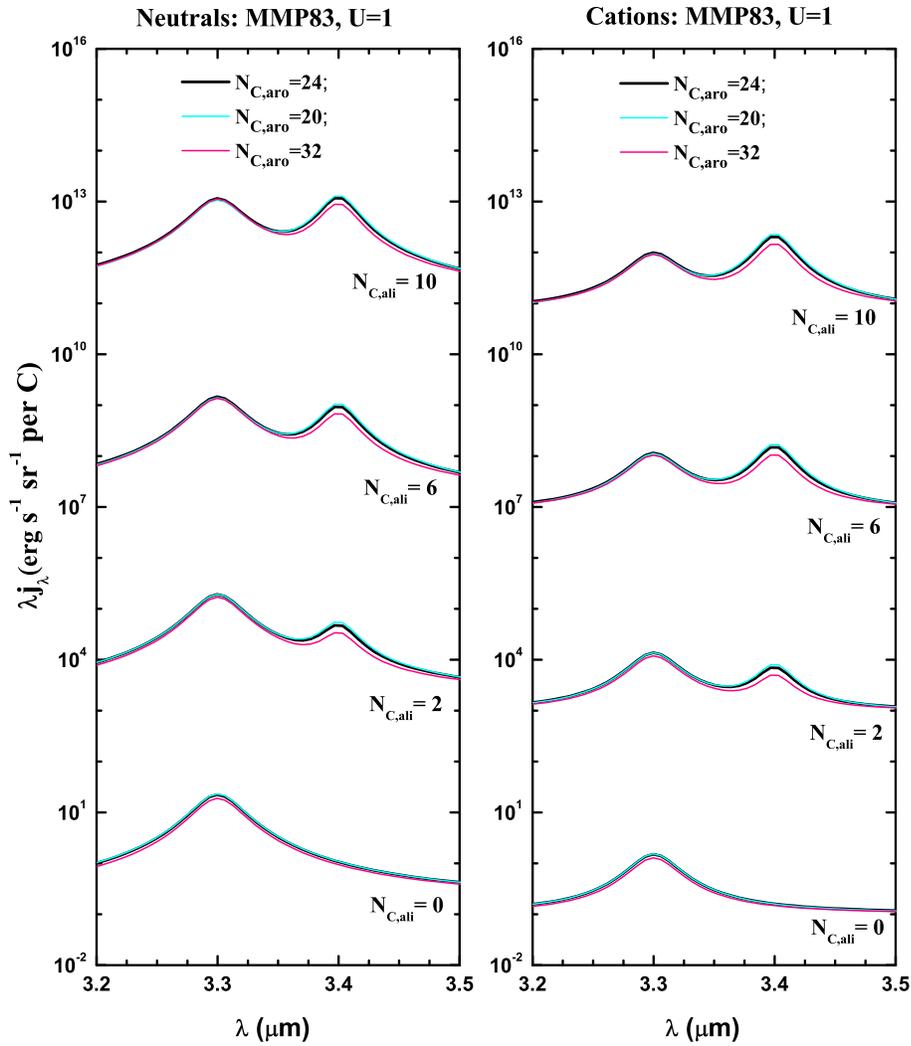}
\vspace{-4cm}
\caption{\footnotesize
              \label{fig:modspec_NC}
              IR emission spectra of neutral (left panel)
              and ionized (right panel) methyl PAHs
              of $\NCali=0, 2, 6, 10$ aliphatic C atoms
              and $\NCaro=20$ (cyan lines),
              $\NCaro=24$ (black lines), or
              $\NCaro=32$ (red lines)
             illuminated by the MMP83 ISRF ($U=1$).
             For clarity, the spectra
             for methyl PAHs with $\NCali=2, 6, 10$
             are vertically shifted.
             }
\end{center}
\end{figure*}

\begin{figure*}
\begin{center}
\includegraphics[scale=0.7,clip]{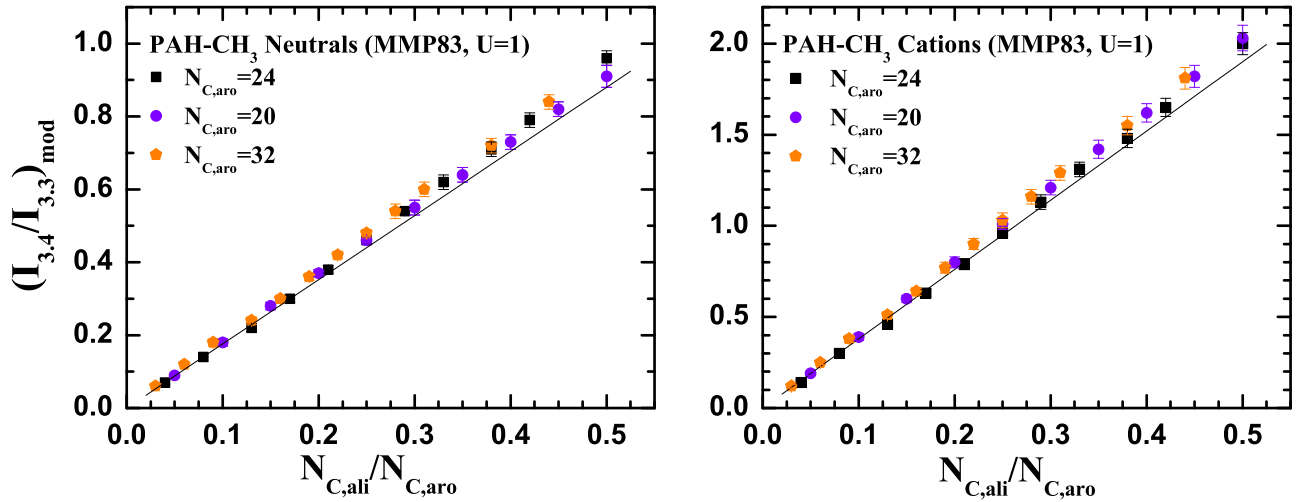}
\caption{\footnotesize
               \label{fig:Iratio_NC}
              Model-calculated intensity ratios $\Iratiomod$
              as a function of the aliphatic fraction $\NCali/\NCaro$
              for neutral methyl PAHs (left panel) of
              $\NCaro=20$ (purple circles),
              $\NCaro=24$ (black squares), and
              $\NCaro=32$ (orange pentagons)
              and their cations (right panel).
              The molecules and their cations are illuminated by
              the MMP83 ISRF ($U=1$).
              The solid black line plots $\Iratiomod =1.76\times
               \left(\NCali/\NCaro\right)$
               for neutrals and $\Iratiomod =3.80\times
               \left(\NCali/\NCaro\right)$
               for cations.
              }
\end{center}
\end{figure*}

So far, we have confined ourselves to
coronene-like PAHs with $\NCaro=24$.
To examine the effects of the PAH size
on the model IR emission spectra
and the band ratios $\Iratiomod$,
we consider methyl PAHs of $\NCaro=20$
aromatic C atoms (like perylene)
and $\NCali=0, 1, 2, ...12$ aliphatic C atoms,
as well as methyl PAHs of $\NCaro=32$
aromatic C atoms (like ovalene)
and $\NCali=0, 1, 2, ...14$
aliphatic C atoms.\footnote{%
   We note that it is not necessary
   to consider larger PAHs
   since the 3.3$\mum$ C--H feature
   is predominantly emitted by
   small neutral PAHs of
   $\simali$20--30 C atoms
   (see Figures~6,7 of Draine \& Li 2007).
   }
As shown in
Figures~\ref{fig:modspec_NC},\ref{fig:Iratio_NC},
neither the IR emission spectra
in the C--H stretch region
nor the band ratios $\Iratiomod$
appreciably differ from each other.

\begin{figure*}
\begin{center}
\includegraphics[scale=0.7,clip]{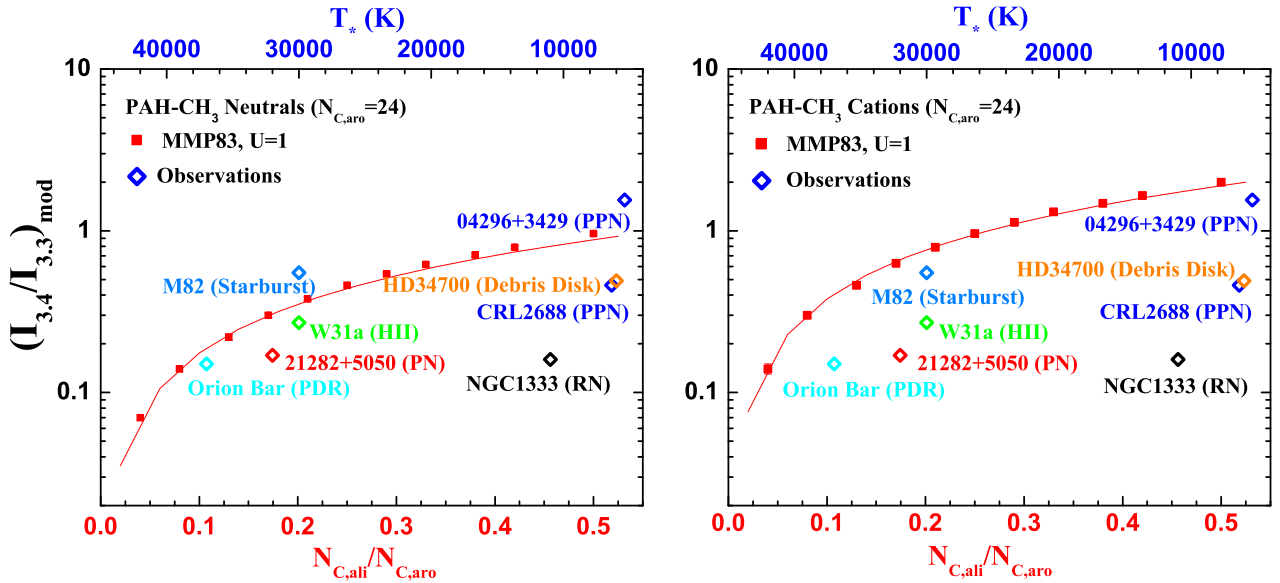}
\caption{\footnotesize
         \label{fig:Iratio_mod}
         Comparison of the band ratios $\Iratioobs$
         observed in the eight representative astrophysical
         environments shown in Figure~\ref{fig:Aro_Ali_Obs}
         with that calculated from neutral (left panel)
         and ionized (right panel) methyl PAHs
         illuminated by the MMP83 ISRF ($U=1$).
         These molecules have $N_{\rm C,aro}=24$ aromatic C atoms
         and a wide range of aliphatic fractions $\NCali/\NCaro$.
         The upper horizontal axis plots the effective temperatures
         of the stars illuminating the observed sources.
         }
\end{center}
\end{figure*}

Finally, we compare in Figure~\ref{fig:Iratio_mod}
the band ratios $\Iratioobs$
observed in the eight representative astrophysical
environments shown in Figure~\ref{fig:Aro_Ali_Obs}
with that calculated from methyl PAHs.
It is seen that the observed band ratios
$\Iratioobs$ of all sources
except the protoplanetary nebula
IRAS~04296+3429 all fall below
the model $\Iratiomod$ curve
of neutral PAHs with $\NCaro=24$
and $\NCali/\NCaro\simlt0.5$.
For IRAS~04296+3429, the unusually
high ratio of $\Iratioobs\approx1.54$
falls below the model $\Iratiomod$ curve
of PAH cations. This demonstrates that a mixture
of neutral and ionized methyl PAHs are
capable of accounting for all the observed
band ratios, including those of protoplanetary
nebulae some of which exhibit an extremely
strong 3.4$\mum$ feature.

\section{Summary}\label{sec:summary}
We have presented an intensity scaling scheme
for scaling the band strengths of the C--H stretching
features of PAHs with a methyl side chain
computed with B3LYP which is less accurate and
computationally less demanding.
Such an intensity scaling approach allows
us to obtain accurate band strengths,
as accurate as that computed with MP2
in conjunction with large basis sets
which is known to be more accurate
than B3LYP but computationally very expensive.
It is found that the band intensities calculated
with {\rm B3LYP/6-31G$^{\ast}$}
for a number of molecules
are much higher than their gas-phase experimental values.
Using better basis sets in conjunction with
the B3LYP method, the computed intensities
are still considerably higher (by $\simali$30\%)
compared to their experimental results.
The MP2 method with the basis set of
{\rm 6-311+G(3df,3pd)} reproduces the measured
      intensities reasonably well.
      However, such calculations are far too expensive
      especially for large molecules.
      It is shown that intensity scaling approaches
      that are based on the B3LYP data
      can be just as successful.

We have also calculated the model spectra
of methylated PAHs and their cations 
of different sizes and various numbers 
of methyl sidegroups,
excited by starlight of different spectral shapes
and intensities.
We find that the ratio of the model intensity of
the 3.4$\mum$ feature
to that of the 3.3$\mum$ feature
is insensitive to the PAH size
and the spectral shape and intensity
of the exciting starlight.
We have derived a simple, convenient, 
and straightforward relation for determining
the aliphatic fraction $\NCali/\NCaro$
of the 3.3$\mum$-band carriers
from the observed band ratios
$\Iratioobs$:
$\NCali/\NCaro\approx0.57\times\Iratioobs$
for neutrals and
$\NCali/\NCaro\approx0.26\times\Iratioobs$
for cations.

\acknowledgments{%
We thank B.T.~Draine, J.Y.~Seok, 
and the anonymous referee
for very helpful suggestions.
AL and XJY are supported in part by
NSFC\,11473023, NSFC\,11273022,
NSF AST-1311804, NNX13AE63G,
Hunan Provincial NSF\,2015JJ3124,
and the University of Missouri Research Board.
RG is supported in part by NSF-PRISM grant
Mathematics and Life Sciences (0928053).
Computations were performed using the high-performance computer
resources of the University of Missouri Bioinformatics Consortium.
}

\appendix
\section{Rationale for A Non-Zero Offset 
           in the Intensity Scaling Relation}
We show here that the non-zero offset 
in the intensity scaling relation 
(see \S\ref{sec:scaling})
comes from the fact that the intensities of
methyl (aliphatic) and aromatic C--H stretches
do not scale alike (i.e., $\fali\neq \faro$).
Eqs.\,\ref{eq:I_L12_define_a}
and \ref{eq:I_L12_define_b}
show the total intensities of
the C--H stretching regions as a function
of the numbers of methyl ($n_{\Ali}$)
and aromatic ($n_{\Aro}$) C--H bonds
and the average IR intensities of a methyl ($\Aali$)
or of an aromatic ($\Aaro$) C--H stretching bond
for two theoretical levels $\Li$ and $\Lj$:
\begin{subequations}\label{eq:I_L12_define}
\begin{align}
A(\Li)&~=~n_{\Ali}\,A_{\Ali}(\Li) + n_{\Aro}\,A_{\Aro}(\Li)
\label{eq:I_L12_define_a} \\
A(\Lj)&~=~n_{\Ali}\,A_{\Ali}(\Lj) + n_{\Aro}\,A_{\Aro}(\Lj)
\label{eq:I_L12_define_b}
\end{align}
\end{subequations}
where $A_{\Ali}(\Li)$ and $A_{\Aro}(\Li)$
are respectively the strengths of
one aliphatic or one aromatic C--H bond
computed at the $\Li$ level,
and $A_{\Ali}(\Lj)$ and $A_{\Aro}(\Lj)$
are the same parameters but computed
at the $\Lj$ level.

Assuming that the intensities of the methyl (aliphatic)
and aromatic C--H stretches scale with factors
$\fali$ and $\faro$, respectively,
one can express the total intensity at level $\Lj$
as a function of the average IR intensities of
a methyl (aliphatic) or of an aromatic
C--H stretching bond at theoretical levels $\Li$
[i.e., $A_{\Ali}(\Li)$ and $A_{\Aro}(\Li)$;
see eq.\,\ref{eq:I_L12_scale_1a}].
By addition and subtraction of
the term $\faro\,n_{\Ali}\,A_{\Ali}(\Li)$,
it is possible to rewrite
eq.\,\ref{eq:I_L12_scale_1a}
such that $A(\Lj)$ is expressed as a function
of $A(\Li)$ and $A_{\Ali}(\Li)$
(see eq.\,\ref{eq:I_L12_scale_1d}).
Using instead the analogous term
$\fali\,n_{\Aro}\,A_{\Aro}(\Li)$ gives $A(\Lj)$
as a function of $A(\Li)$ and $A_{\Aro}(\Li)$
(see eq.\,\ref{eq:I_L12_scale_2d}).
\begin{subequations}\label{eq:I_L12_scale1}
\begin{align}
A(\Lj)&~=~\fali\,n_{\Ali}\,A_{\Ali}(\Li) + \faro\,n_{\Aro}\,A_{\Aro}(\Li)
\label{eq:I_L12_scale_1a} \\
     &~=~\fali\,n_{\Ali}\,A_{\Ali}(\Li) + \faro\,n_{\Aro}\,A_{\Aro}(\Li)
       + \faro\,n_{\Ali}\,A_{\Ali}(\Li) - \faro\,n_{\Ali}\,A_{\Ali}(\Li)
\label{eq:I_L12_scale_1b} \\
     &~=~\faro\,[n_{\Ali}\,A_{\Ali}(\Li) + n_{\Aro}\,A_{\Aro}(\Li)]
       + \fali\,n_{\Ali}\,A_{\Ali}(\Li) - \faro\,n_{\Ali}\,A_{\Ali}(\Li)
\label{eq:I_L12_scale_1c} \\
     &~=~\faro\,A(\Li) + \underline{(\fali - \faro)\,n_{\Ali}\,A_{\Ali}(\Li)}
\label{eq:I_L12_scale_1d}
\end{align}
\end{subequations}
or
\begin{subequations}\label{eq:I_L12_scale2}
\begin{align}
A(\Lj)&~=~\fali\,n_{\Ali}\,A_{\Ali}(\Li) + \faro\,n_{\Aro}\,A_{\Aro}(\Li)
\label{eq:I_L12_scale_2a} \\
     &~=~\fali\,n_{\Ali}\,A_{\Ali}(\Li) + \faro\,n_{\Aro}\,A_{\Aro}(\Li)
       + \fali\,n_{\Aro}\,A_{\Aro}(\Li) - \fali\,n_{\Aro}\,A_{\Aro}(\Li)
\label{eq:I_L12_scale_2b} \\
     &~=~\faro\,[n_{\Ali}\,A_{\Ali}(\Li) + n_{\Aro}\,A_{\Aro}(\Li)]
       + \faro\,n_{\Aro}\,A_{\Aro}(\Li) - \fali\,n_{\Aro}\,A_{\Aro}(\Li)
\label{eq:I_L12_scale_2c} \\
     &~=~\fali\,A(\Li) + \underline{(\faro - \fali)\,n_{\Aro}\,A_{\Aro}(\Li)}
\label{eq:I_L12_scale_2d}
\end{align}
\end{subequations}
where the underlined terms in
eqs.\,\ref{eq:I_L12_scale_1d}
and \ref{eq:I_L12_scale_2d}
are responsible for the offset
in the correlations between the total intensities
at levels $\Li$ and $\Lj$,
and these offsets vanish only when $\faro = \fali$.
This condition never holds
and, in addition, it also is not trivial to determine
at what level $\faro$ and $\fali$ converge.
We have extensively studied
the basis set effects
at the B3LYP level for toluene
and the three isomers of methylpyrene
(see Yang et al.\ 2013).
There is a very large basis set dependency
in that $A_{\Aro}$ is greatly
reduced with the improvements of the basis set.
The typical $A_{\Aro}$ value
at the {\rm B3LYP/6-31G$^{\ast}$} level
is $\simali$18--20$\km\mol^{-1}$
and this value drops to $\simali$12.5--13.3$\km\mol^{-1}$
at the highest level {\rm 6-311++G(3df,3pd)},
i.e., a scaling factor of $\faro\approx0.7$.
In contrast, the basis set dependency
of $A_{\Ali}$ is less than that of $A_{\Aro}$.
A typical $A_{\Ali}$ value
at the {\rm B3LYP/6-31G$^{\ast}$} level
is $\simali$23--27$\km\mol^{-1}$ and
this value drops to $\simali$19--24$\km\mol^{-1}$
at the {\rm 6-311++G(3df,3pd)} level,
i.e., a scaling factor of $\fali\approx0.85$.
This confirms the need for non-zero offset
in intensity scaling because $\fali\neq \faro$.




\end{document}